\documentclass[prl,aps,amsmath,amssymb,twocolumn,superscriptaddress]{revtex4-2}

\usepackage{graphicx}
\usepackage{dcolumn}
\usepackage{bm}
\usepackage{graphicx}%
\usepackage{multirow}%
\usepackage{amsmath,amssymb,amsfonts}%
\usepackage{amsthm}%
\usepackage{mathrsfs}%
\usepackage[title]{appendix}%
\usepackage{xcolor}%
\usepackage{textcomp}%
\usepackage{booktabs}%
\usepackage{algorithm}%
\usepackage{algorithmicx}%
\usepackage{algpseudocode}%
\usepackage{listings}%
\usepackage{times}
\usepackage{amssymb}
\usepackage{color}
\usepackage{amsmath}
\usepackage{amsbsy}
\usepackage{amsthm}
\usepackage{bbm}
\usepackage{epsfig}
\usepackage{float}
\usepackage{subfigure}
\usepackage{hyperref}

\begin{document}


\title{Observation of gyroscopic coupling in a non-spinning levitated ferromagnet}

\author{Felix Ahrens}
\affiliation{Istituto di Fotonica e Nanotecnologie IFN-CNR, 38123 Povo, Trento, Italy}
\affiliation{Fondazione Bruno Kessler, 38123 Povo, Trento, Italy}

\author{Andrea Vinante}
\email{andrea.vinante@cnr.it}
\affiliation{Istituto di Fotonica e Nanotecnologie IFN-CNR, 38123 Povo, Trento, Italy}
\affiliation{Fondazione Bruno Kessler, 38123 Povo, Trento, Italy}

\date{\today}

\begin{abstract}
A non-spinning permanent ferromagnet is predicted to behave as a gyroscope at sufficiently low frequencies, which can be seen as a manifestation of gyromagnetism. This yet unexplored regime, conjectured for the first time by Maxwell in 1861, has recently been proposed for ultrasensitive magnetometry and for atomic-like quantum stabilization of a levitated nanomagnet in a static field. Here, we observe signatures of gyroscopic effects in the rotational dynamics of a non-spinning permanent ferromagnet levitated in a superconducting trap. Specifically, we detect spin-rotation coupling between different librational modes, leading to elliptical trajectories, in good agreement with theoretical predictions. From our measurements, we can infer both the intrinsic angular momentum of the levitated magnet and its gyromagnetic $g$-factor.
\end{abstract}

\maketitle

The concept of spin is unique to quantum mechanics and has no classical interpretation. Nevertheless, it is an established experimental fact that quantum spin corresponds to a genuine form of mechanical angular momentum. This relationship, known as gyromagnetism, was demonstrated by the famous Einstein-de Haas \cite{Einstein1915} and Barnett \cite{Barnett1915} experiments, which showed that changes in the magnetization of a ferromagnet are associated to changes in its rotational state and vice versa. Such effects are typically observed in soft magnetizable materials, where the internal rearrangement of magnetic dipoles occurs
via spin-lattice relaxation \cite{Dornes2019}. It is worth noting, however, that the first attempt to detect gyromagnetic effects was performed much earlier, namely by Maxwell in 1861, and involved a simpler idea: observing the gyroscopic behavior of a whole piece of magnetized material \cite{Maxwell}. Despite Maxwell's brilliant intuition, his experiment was unsuccessful, and was later regarded extremely challenging \cite{Barnett1935}. 

Recent proposals have revived Maxwell's idea by predicting that the angular momentum arising from fixed magnetization in a hard ferromagnet should lead to gyroscopic dynamics, even if the ferromagnet is not spinning at all \cite{Kimball2016, Rusconi2017, Rusconi2017b, Fadeev2021gravity, Fadeev2021,Ni2025}. 
For sufficiently slow rotational motion and weak torque, the ferromagnet would thus exhibit precessional dynamics akin to a spinning top. This regime can be observed only if the induced precession frequency $\Omega$ is lower than the critical value:
\begin{equation}
\Omega \ll \omega_I = S/I
\end{equation}
where $S$ is the intrinsic angular momentum and $I$ is the moment of inertia, that is, when the precessional angular momentum $L=I\Omega$ is lower than the intrinsic angular momentum $S$. The quantity $\omega_I$ is the so-called Einstein-de Haas frequency of the ferromagnet. For a sphere, the scaling with the radius $R$ is $\omega_I \propto 1/R^2$, which explains why this effect can hardly be observed in macroscopic systems. If the torque is induced by a weak external magnetic field $B$, the ferromagnet will behave as a giant atomic spin, featuring Larmor precession with frequency $\Omega_\mathrm{L} = \gamma_0 B$, where $\gamma_0$ is the mean gyromagnetic ratio of the spins \cite{Kimball2016,Ni2025}. Conversely, if the Larmor precession is higher than the Einstein-de Haas frequency $\Omega_\mathrm{L} > \omega_I$, the ferromagnet will exhibit classical compass-like librational behavior \cite{Vinante2021,Kalia2024}. 

The gyroscopic regime may enable ultrasensitive dc magnetometry in a way similar to atomic spins, with the potential to overcome the standard quantum limits on atomic magnetometers by many orders of magnitude \cite{Kimball2016,Fadeev2021,Ni2025}. Furthermore, it has been proposed for testing the relativistic frame drag with quantum spin \cite{Fadeev2021gravity}. Remarkably, a levitated ferromagnet can exhibit extreme magnetic field resolution even in the standard librational regime, as in a recent ac magnetometry demonstration \cite{ahrens2025}.
A different application of the gyroscopic nature of a ferromagnet has been proposed \cite{Rusconi2017, Rusconi2017b}, namely a nanomagnet could be stabilized in an external static field, such as in an Ioffe-Pritchard trap, similarly to an atomic spin. This would represent a violation of the classical Earnshaw theorem \cite{Bassani2006}.

\begin{figure}[!ht]
\includegraphics[width=\columnwidth]{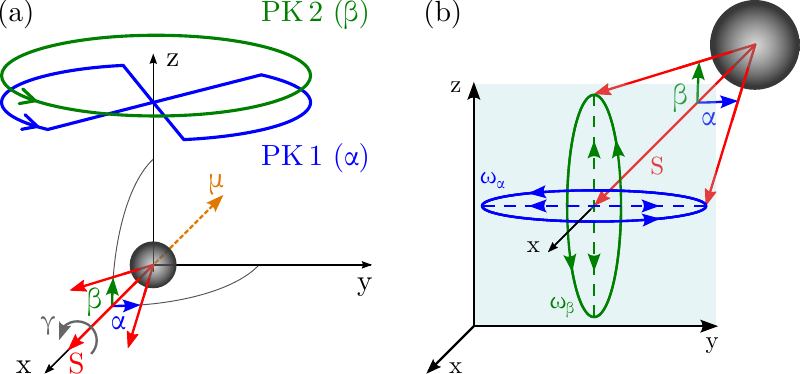} 
\caption{
(a) Model and conventions adopted. A ferromagnet with spin $\bm S$ and magnetic moment $\bm \mu=-\gamma_0 \bm S$ is trapped along the $x$ axis. It can harmonically librate around $z$ by an angle $\alpha$ and around $y$ by an angle $\beta$. Two superconducting pick-up coils coupled to a dc SQUID (not shown) are placed above the ferromagnet. Coil PK1 is an 8-shape coil optimized for detection of $\alpha$ motion and ideally insensitive to $\beta$ motion. Coil PK2 is a circular coil sensitive to $\beta$ and ideally insensitive to $\alpha$.
(b) The elliptical motion of the quasi-$\alpha$ (quasi-$\beta$) modes with rotation frequency $\omega_\alpha$ ($\omega_\beta$) arising from gyroscopic effects is highlighted in blue (green). The hypothetical purely librational $\alpha$ and $\beta$ modes in the absence of gyroscopic effects are indicated as dotted lines.} \label{fig1}
\end{figure}

Here, we present an experiment that directly probes gyromagnetic effects in a macroscopic permanent hard ferromagnet. In our experiment the dominant dynamics is still librational, i.e.~non-gyroscopic. Nevertheless we can infer gyroscopic spin-rotation coupling between the librational modes as a second order effect by observing elliptical trajectories in the angular motion, as depicted in Fig.\,\ref{fig1}b.

Our apparatus consists of a hard ferromagnetic microsphere based on a rare earth alloy \cite{Magnequench}, with radius in the range $20-30 \,\mu$m, levitated by the Meissner effect in a cylindrically symmetrical superconducting trap based on a type-I superconductor at temperature $T=4.18$ K. This setup has been modeled in several previous papers based on the image method \cite{Vinante2020, Vinante2021, Vinante2022, ahrens2025}. 
The key aspects for our experiment are depicted in Fig.\,\ref{fig1}a. We define $\bm{\mu}=-\gamma_0 \bm{S}$ as the magnetic dipole moment, where $\bm{S}$ is the total spin angular momentum and $\gamma_0$ is the gyromagnetic ratio.
At equilibrium $\bm \mu$ lies approximately on the horizontal plane. For convenience, let us assume the orientation of $\bm \mu$ in the $x$ direction as in Fig.\,\ref{fig1}a. The rotational dynamics of the magnet features two trapped librational modes with very different resonance frequency around the $z$ axis ($\alpha$ mode) and the $y$ axis ($\beta$ mode). Trapping is provided by a combination of Meissner field (Fig.\,\ref{fig3}b) and a residual field. We discuss the physical mechanisms in the End Matter section. 
We can define the harmonic restoring torque for small $\alpha$ and $\beta$ librations as $T_z=-I \omega_\alpha^2 \alpha$ and $T_y=I \omega_\beta^2 \beta$. Here, $I$ is the isotropic moment of inertia (we will later discuss the possible impact of anisotropies), and $\omega_\alpha$, $\omega_\beta$ are resonance frequencies of the librational modes. Experimentally, frequencies are about $f_\alpha =\omega_\alpha/(2\pi) \sim 100$\,Hz and $f_\beta =\omega_\beta/(2\pi) \sim 400$\,Hz.


To assess the contribution of gyroscopic effects, we need to solve the most general equation of motion of a ferromagnetic rotor \cite{Fadeev2021, Vinante2021}:
\begin{align} 
\begin{split}
\dot {\bm J} &= \bm {T}  \\
\dot{\bm S} &= \bm \Omega \times \bm S   \label{Jdot}
\end{split}
\end{align}
Here, $\bm J = \bm L + \bm S$ is the total angular momentum, with $\bm L = I \bm \Omega$ being the kinetic angular momentum, and $\bm S$ is the intrinsic angular momentum (which may include both spin and orbital atomic angular momentum), $\bm T$ is the torque. Although the latter could include any additional torque, here we consider only the restoring torque previously discussed. Note that the first of Eqs.\,(\ref{Jdot}) states the conservation of total angular momentum, while the second arises from the assumption that the intrinsic spin is rigidly attached to the magnet crystal, i.e.~$\bm S = S \bm {\hat n}$ where $\bm {\hat n}$ is the unit vector along the easy axis of the magnet. This is a good assumption for a hard ferromagnet, since the internal magnetization dynamics, with characteristic frequencies at several gigahertz, is up to ten orders of magnitude faster than the mechanical motion \cite{Kimball2016}. 

The equations can be solved in a general framework \cite{Kalia2024}, but for our purposes it is sufficient to find the solution in the linear approximation for small angles $\alpha$,\,$\beta$ \cite{Vinante2021}. To this end, we write the total spin as $\bm S = S \left( 1,\alpha,\beta \right)$ and the angular velocity vector as $\bm \Omega\simeq \left(0, -\dot \beta, \dot \alpha \right)$. Inserting in Eqs.\,(\ref{Jdot}) and retaining only linear terms in $\alpha$ and $\beta$ we obtain: 
\begin{align} 
\begin{split} \label{eqalphabeta}
\ddot \alpha &+  \omega_\alpha^2 \alpha  + \omega_I \dot \beta = 0    \\
\ddot \beta &+ \omega_\beta^2 \beta -\omega_I \dot \alpha =0        
\end{split}
\end{align}
In the latter equations, we have neglected  dissipation. Although the latter is always present in real experiments, we performed measurements over a time shorter than the damping time. We have also neglected the classical rotations around the $\bm S$ vector, i.e.~we have assumed $\dot \gamma = 0$, where $\gamma$ is the rotation angle (see Fig.\,\ref{fig1}a). We will discuss later the validity of this assumption. 

Equations (\ref{eqalphabeta}) define coupled linear harmonic oscillators with kinetic coupling terms.
In the case of well-separated $\omega_\alpha$ and $\omega_\beta$, relevant to our experiment, the solutions of Eqs.\,(\ref{eqalphabeta}) describe elliptical oscillations with a dominant axis and a secondary axis, as shown in Fig.\,\ref{fig1}b. In particular, we define a quasi-$\alpha$ mode, with $\alpha$ and $\beta$ oscillating according to:
\begin{align} 
\begin{split} \label{quasialpha}
&\alpha (t) = \alpha_0 \mathrm{sin} (\omega_\alpha t) ,   \\
&\beta (t) = g_\alpha \alpha_0 \mathrm{cos} (\omega_\alpha t) , \\
&g_\alpha = \frac{\omega_\alpha \omega_I}{\omega_\beta^2-\omega_\alpha^2}
\end{split}
\end{align}
and similarly a quasi-$\beta$ mode, with $\alpha$ and $\beta$ oscillating according to:
\begin{align} 
\begin{split} \label{quasibeta}
&\beta (t) = \beta_0 \mathrm{sin} (\omega_\beta t) , \\
&\alpha (t) = g_\beta \beta_0 \mathrm{cos} (\omega_\beta t). \\
&g_\beta = \frac{\omega_\beta \omega_I}{\omega_\beta^2-\omega_\alpha^2}
\end{split}
\end{align}
Thus, the $\pi/2$ phase shift arising from the gyromagnetic coupling in Eqs.\,(\ref{eqalphabeta}) implies elliptical trajectories in the $\alpha-\beta$ plane. The amplitudes on the secondary axes are suppressed by factors $g_{\alpha,\beta}$ that satisfy $g_{\alpha,\beta}\ll 1$ as long as $\omega_I \ll \omega_\alpha , \omega_\beta$, a condition that holds in our experiment.

In order to experimentally detect the elliptical trajectories expressed by Eqs.\,(\ref{quasialpha},\ref{quasibeta}) we need to monitor the motion through at least two independent channels. To this end, in our setup we implement two detection channels with different sensitivity to $\alpha$ or $\beta$ motion (labeled as $1$ and $2$) based on two different superconducting pick-up coils (PK1 and PK2) depicted in Fig.\,\ref{fig1}a. The coils, connected to two independent Superconducting Quantum Interference Devices (SQUIDs), measure the magnetic flux caused by the magnet motion. Both coils lay in the same horizontal plane about $2$\,mm above the magnet and are centered on the trap symmetry axis. The coil PK1 is designed as an 8-shape so that it is coupled to $\alpha$ motion and ideally insensitive to $\beta$, while PK2 is designed as a circular coil so that it is coupled to $\beta$ and is ideally insensitive to $\alpha$. In practice we achieve an imperfect selectivity, and each coil is typically coupled to the other degree of freedom with a suppression factor of the order of a few percent. 

\begin{figure}[!ht]
\includegraphics[width=8.6cm]{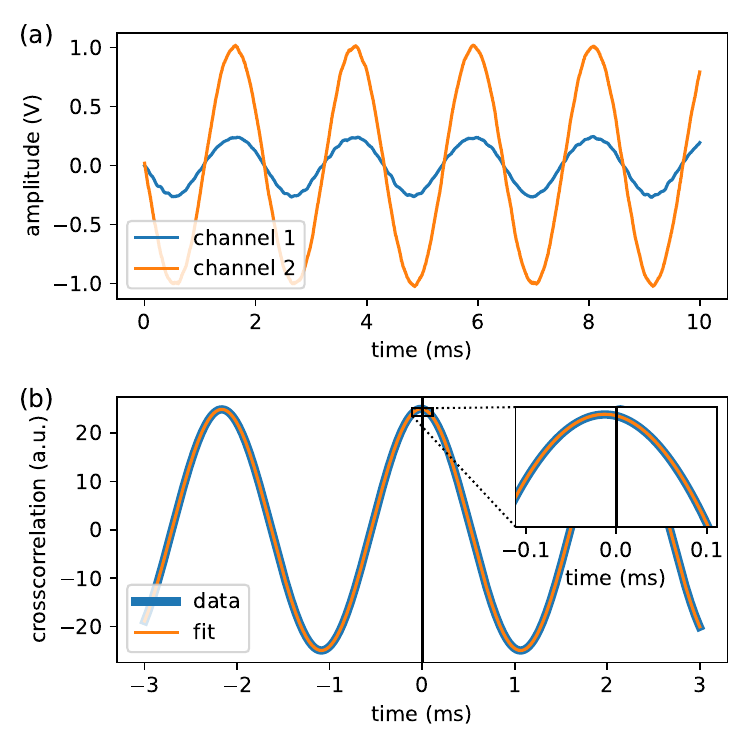} 
\caption{Representative samples of raw and cross-correlation data.
(a) Sample of raw data from channel 1 and channel 2 with quasi-$\beta$ mode excited, over a short time of 10\,ms. The frequency of the quasi-$\beta$ mode is $4
64.4$\,Hz. (b) Cross-correlation $C_{12}$ calculated with the same setting of (a) with a total averaging time of 500\,ms. In the inset, the cross-correlation is zoomed close to zero time lag. The phase shift with respect to the pure cosine behavior reveals an out-of-phase component as expected from Eqs.\,(\ref{quasibeta}).}  \label{fig2}
\end{figure}


We take imperfections into account by writing the SQUID output voltage in the channels $1$ and $2$ due to the $\alpha$ and $\beta$ motion as: 
\begin{align}
\begin{split}  \label{channels}
&V_1 \left( t \right)= A \alpha \left( t \right)+ B \beta \left( t \right) \\
&V_2 \left( t \right) = C \alpha \left( t \right) + D \beta \left( t \right)    
\end{split}
\end{align}
with the only assumption that by design $|A| \gg |B|$ and $|D| \gg |C|$.
The presence of the cross-coupling factor $C$ implies that when the quasi-$\alpha$ mode is excited, two components will be detected by the secondary channel $2$, one in-phase with the main channel $1$ due to the cross-coupling $C$, and one out-of-phase due to the genuine $\beta$ motion arising from gyromagnetic coupling through the coupling $D$. Specular considerations hold for the quasi-$\beta$ mode.
As a representative example, we show in Fig.\,\ref{fig2}a a sample of raw data from channel 1 and 2 with only the quasi-$\beta$ mode excited, for a microsphere of radius 23.6\,\textmu m. At first glance, the two signals are in phase and it is difficult to distinguish the out-of-phase component predicted by Eqs.\,(\ref{quasibeta}).

In order to unambiguously distinguish an out-of-phase component in the presence of noise, we implement a cross-correlation technique. We define the correlation functions as:
\begin{equation}
C_{ij}\left( \tau \right) =\int_{0}^{t_\mathrm{m}} V_i \left( t \right) V_j \left( t + \tau \right) \mathrm{d}t  \label{Cij}
\end{equation}
where $\tau$ is a time lag and the averaging period $t_\mathrm{m}$ is much longer than the oscillation period. Experimentally, the integral is replaced by a sum over the sampled values. We consider two quasi-stationary cases, in which we excite either the quasi-$\alpha$ mode or the quasi-$\beta$ modes to an amplitude much higher than the noise level. For each case the correlation functions are of the form:
\begin{align}
\begin{split}
\label{correlation}
&C_{ij}^{(\alpha)} \left( \tau \right)=  c_{ij}^{(\alpha)} \mathrm{cos} \left( \omega_{\alpha}\tau \right) + s_{ij}^{(\alpha)} \mathrm{sin} \left( \omega_{\alpha} \tau \right)  \\
&C_{ij}^{(\beta)} \left( \tau \right)= c_{ij}^{(\beta)} \mathrm{cos} \left( \omega_{\beta} \tau \right) + s_{ij}^{(\beta)} \mathrm{sin} \left( \omega_{\beta} \tau \right) 
\end{split}
\end{align}
We extract the in-phase and out-of-phase components $c_{ij}$, $s_{ij}$ by fitting the experimentally measured correlation functions. Details of the actual numerical procedure are reported in the Supplemental Information \cite{Supplementary}. Gyroscopic coupling appears as a non-zero value of the out-of-phase terms $s_{12}^{(\alpha)}$ and $s_{21}^{(\beta)}$ while crosstalk through the stray couplings $B$ and $C$ will appear in the $c_{12}$ terms.
As an example, Fig.\,\ref{fig2}b shows the cross-correlation $C_{12}$ for the same setting of Fig.\,\ref{fig2}a evaluated over an averaging time $t_\mathrm{m}=500$ ms. Here, a phase shift is clearly visible in the data, revealing a finite out-of-phase component $s_{12}$. 

In order to obtain quantitative information on the spin-rotation coupling expressed by the Einstein-de Haas frequency $\omega_I$, we combine correlation measurements on the quasi-$\alpha$ mode and the quasi-$\beta$ mode. By means of Eqs.\,(\ref{quasialpha}, \ref{quasibeta}, \ref{channels}, \ref{Cij}, \ref{correlation}) and under the assumptions of partial selectivity $|A| \gg |B|$ and $|D| \gg |C|$ and with the further assumption that $\omega_I \ll \omega_\alpha , \omega_\beta$, one can show that:
\begin{equation}
  r_\alpha r_\beta = \frac{\omega_\alpha \omega_\beta}{\left( \omega_\beta^2-\omega_\alpha^2 \right)^2} \omega_I^2   \label{fI}
\end{equation}
where we have defined:
\begin{align}
\begin{split}
r_\alpha = \frac{s_{12}^{(\alpha)}}{c_{11}^{(\alpha)}}\\
r_\beta =  \frac{s_{21}^{(\beta)}}{c_{22}^{(\beta)}} 
\end{split}
\end{align}

The notable feature of Eq.\,(\ref{fI}) is that it directly connects an experimentally measurable quantity (left-hand side) with a combination of the mode frequencies and the unknown Einstein-de Haas frequency $\omega_I$, while it does not contain any of the coupling constants $A$, $B$, $C$ and $D$. Therefore, it provides a calibration-free procedure for extracting $\omega_I$ from the data. Experimentally, we evaluate $r_\alpha$ and $r_\beta$ as mean values over many repetitions of the cross-correlation measurement, while their uncertainty is estimated with the mean standard error.

We have performed measurements with four experimental configurations, summarized in Table\,\ref{table}. We have used three different microspheres with the same material and different radius, and for one of them we have performed two different cooldowns resulting in a very different $\alpha$ mode frequency. For each setting, we have measured the Einstein-de Haas frequency $f_I=\omega_I/(2\pi)$ using the procedure above, which is shown in the third column of Table\,\ref{table}. Note that this measurement is calibration-free, so it provides an absolute estimate of the intrinsic angular momentum $S$ normalized to the moment of inertia $I$.
For each particle, we determine the radius $R$ and magnetization $M$ from the observed mode frequencies. In fact, according to the image method, the frequency modes $\beta$ and $z$ are uniquely determined by $R$ and $M$, as long as the mass density $\rho$ and the gravity acceleration $g_0$ are fixed \cite{ahrens2025}. This procedure is described in detail in the Supplemental Information \cite{Supplementary}. From $R$, $M$ and $\rho$, we can derive the mass $m$, the magnetic moment $\mu$ and the moment of inertia $I$. The inferred values of $R$ and $M$ in the experiment are reported in Table\,\ref{table}. In particular, the inferred values of $R$ are consistent with less precise estimations based on direct optical inspection with a low-magnification microscope during the particle loading process in the trap.

 \begin{table}
\vspace{0.4cm}
\begin{tabular}{|c ||c|c|c|  }
 \hline
 
  $R$ (\textmu m)& $M$ (kA/m) & $f_I$ (Hz) & $g$ \\ 
 \hline
 $31.2 \pm 0.4$ & $591 \pm 18$ & $0.33\pm 0.04$ &  $1.11 \pm 0.14$ \\
 \hline
 $23.6 \pm 0.2$ & $675 \pm 20$ & $0.62 \pm 0.02$ & $1.19 \pm 0.04$\\
 \hline
 $19.0 \pm 0.2$ &  $574 \pm 17$  & $0.88 \pm 0.05$   & $1.10 \pm 0.07$\\
 \hline
 $18.8 \pm 0.2$ & $581 \pm 16$ & $0.86\pm 0.03$ &  $1.16 \pm 0.04$ \\
 \hline
 \end{tabular}
\caption{Experimental values inferred from the different measurements with different particles. The gyromagnetic frequency $f_I$ is estimated using the procedure described in the text. The radius $R$ and magnetization $M$ are inferred from the observed resonant frequencies of $\beta$ and $z$ mode as described in the Supplemental Information. From $R$ and $M$ we obtain the mass $m$ and the magnetic moment $\mu$. The $g$ factor (4th column) is derived from $\mu$ and $f_I$. Note that the 3rd and 4th rows are obtained from measurements on the same particle in two different cooldowns, with significantly different $\alpha$ mode frequencies.}  \label{table}
\end{table}

Finally, we can estimate the gyromagnetic $g-$factor from the definition:
\begin{equation}
    g = \frac{\mu/\mu_\mathrm{B}}{S/\hbar} 
\end{equation}
where $\mu_\mathrm{B}$ is the Bohr magneton and $\hbar$ is the reduced Planck constant. The values of $g$ inferred from the data, reported in the last column of Table\,\ref{table}, are consistent with each other, in agreement with the fact that measurements have been performed with particles made of the same magnetic material. 

The numerical values of $g$ agree reasonably well with the expectations for a rare-earth magnet. The alloy used in this experiment is based on a commercial Nd-Pr-Fe-Co-B alloy. We used this alloy because it is known to maintain magnetization down to cryogenic temperatures, unlike standard NdFeB which may undergo phase transitions. We can reasonably assume that the basic structure is Nd$_2$Fe$_{14}$B with Nd partially replaced by Pr for the rare-earth component, and Fe partially replaced by Co. Taking as reference Nd$_2$Fe$_{14}$B, we estimate a mean effective value for $g$:
\begin{equation}
  g_{\mathrm{eff}}=\frac{2 g_\mathrm{Nd} S_\mathrm{Nd} + 14 g_\mathrm{Fe} S_\mathrm{Fe}}{2 S_\mathrm{Nd} +14 S_\mathrm{Fe} } = 1.28
\end{equation}
where we have used the literature values $g_\mathrm{Nd}=\frac{8}{11}$, $S_\mathrm{Nd}=\frac{9}{2} \hbar$ for the Nd$^{3+}$ ion and $g_\mathrm{Fe}=2$, $S_\mathrm{Fe}=\frac{1}{2} \hbar$ for Fe \cite{Buschow}. Note again that with $S$ we mean here the total intrinsic angular momentum, which for rare-earth ions includes both spin or orbital atomic angular momentum. A value of only slightly higher $1.36$ is obtained by replacing Nd with Pr. Thus, our experimental estimations are about $10 \%$ off the simplest prediction for standard hard ferromagnets based on Nd or Pr.

In spite of the substantial agreement of the experimental data with our spin-rotation coupling model, we need to discuss other mechanisms that could potentially mimic the same results. As shown in the Supplemental Information \cite{Supplementary}, a classical angular velocity $\dot \gamma$ around the vector $\bm S$ would play the same role as the Einstein-de Haas frequency $\omega_I$. This is entirely expected on the basis of the Einstein-de Haas equivalence of intrinsic and rotational angular momentum. The coupling between librational modes mediated by $\gamma$ has been indeed observed recently in levitated nanoparticles \cite{Zielinska2023}. Our SQUID-based detection is blind to $\gamma$ motion, but we can reasonably assume that it is dominated by thermal noise, since in our setup both librational modes $\alpha$ and $\beta$ are thermal \cite{ahrens2025,Supplementary}. In this case, $\dot \gamma$ is a process with zero mean and standard deviation $\dot \gamma_\text{rms} = \sqrt{k_\mathrm{B} T/I}$. For all experimental configurations in Table\,\ref{table}, we estimate $\dot \gamma_{\text{rms}} <0.01 \omega_I$. Thus, the effect of $\dot \gamma$ is practically negligible, although it can affect the experimental fluctuations of $\omega_I$. Another possible mechanism that could generate a spin-rotation coupling is a deviation from the perfect spherical shape of the particle, implying an anisotropic moment of inertia. However, in this case, the additional mixing between $\alpha$ and $\beta$ would not be kinetic, so it would not lead to out-of-phase motion \cite{Supplementary}. A relevant coupling could be introduced in combination with gas damping, but is estimated negligible \cite{Martinetz2023}. Finally, we ensured suppression of any instrumental source of differential delay between the two channels \cite{Supplementary}.




In conclusion, we have observed elliptical trajectories in the librational oscillations of an angularly confined levitated hard ferromagnet, 
in qualitative agreement with the theoretical model. In particular, the inferred angular momentum and magnetic moment, encoded in the Einstein-de Haas frequency $f_I$ and the gyromagnetic factor $g$ factor, agree well with expectations for a rare-earth ferromagnet. To our knowledge, this is the first observation of gyromagnetic coupling in the dynamics of a non-spinning permanent magnet, an effect conjectured by Maxwell more than 160 years ago \cite{Maxwell}.

In the future, full gyroscopic dynamics should be accessible. For a superconducting trap, a necessary condition is $\omega_I \gg \omega_{\alpha},\omega_{\beta}$ which requires nanomagnets with radius $R<500$\,nm. A partial gyroscopic regime where $\omega_\beta \gg \omega_{I} \gg \omega_{\alpha}$, which would exhibit a modified Larmor precession \cite{Fadeev2021, Kalia2024} is, in principle, achievable in the present setup by a proper external compensation of stray fields causing the $\alpha$ trapping. However, gyroscopic dynamics of a non-spinning ferromagnet is a general effect, which can be observed using other platforms, such as Ioffe-Pritchard traps \cite{Rusconi2017, Rusconi2017b}, Paul traps \cite{Perdriat2023,Dania2024}, levitated soft ferromagnets \cite{Fuwa2023} or with magnets in free fall \cite{Fadeev2021gravity}. Besides ferromagnets, even the spin-rotation coupling associated to a single individual spin can become an important effect in sufficiently small levitated nanoobjects, which suggests its potential use to perform quantum experiments with nanorotors \cite{Ma2021,Wachter2025}.  

The data that support the findings of this article are openly available \cite{dataset}.

\begin{acknowledgments}
AV thanks Ben Stickler, Andrejs Cebers and Dmitry Budker for useful discussions. We acknowledge support from the QuantERA II Programme (project LEMAQUME) that has received funding from the European Union’s Horizon 2020 research and innovation programme under Grant Agreement No 101017733, and from the Italian Ministry for University and Research within the Italy-Singapore Scientific and Technological Cooperation Agreement 2023-2025. 
\end{acknowledgments}

\bibliography{reference}

\begin{appendices}

\section{END MATTER}

In this section, we will discuss the physical mechanisms that lead to the angular confinement of the magnetic dipole in the two angles $\alpha$ and $\beta$ with well-separated resonance frequencies. This is important to support the validity of the model in Fig.\,\ref{fig1}. The ideal case of a magnetic dipole levitated above a superconducting plane, or a spherical surface as (approximately) in our experiment, can be solved analytically using the image method \cite{Vinante2020, Vinante2021, Vinante2022, ahrens2025}. Analytical details for the spherical case are reported in the Supplemental Information \cite{Supplementary}.

\begin{figure}[!h]
\includegraphics[width=8.6cm]{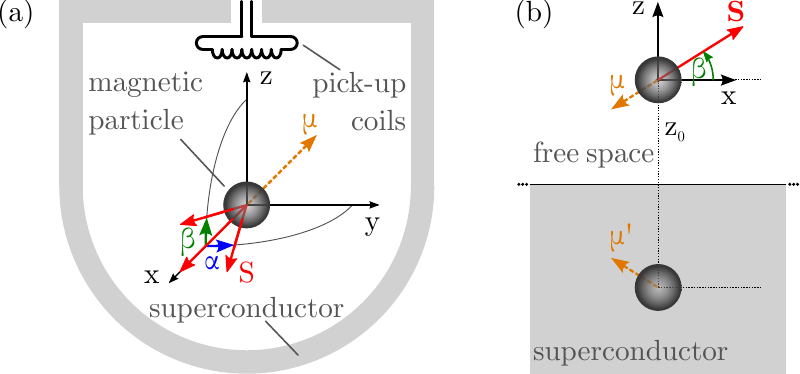} 
\caption{(a) Schematics of the magnetic particle levitated inside a hemispherical superconducting trap. (b) Scheme of a ferromagnet above an infinite superconducting plane (gray region) in the Meissner regime. In this implementation, harmonic confinement of the $\beta$ angle arises naturally due to the interaction of the magnet dipole $\bm \mu$  with an image dipole $\bm \mu'$.}  \label{fig3}
\end{figure}

With reference to Fig.\,\ref{fig3}b, the model provides an analytical expression of the magnetic interaction potential between magnet and superconductor and, by adding the effect of gravity, an analytical determination of the equilibrium configuration, the frequency $f_z$ of the translational $z$-mode, and the frequency $f_\beta$ of the librational mode on the polar angle $\beta$. At equilibrium, the dipole is predicted to lie horizontally. If the system features cylindrical symmetry, the potential is invariant under rotations about the azimuthal angle $\alpha$, which is therefore not confined. Similarly, the rotation about $\gamma$ is not confined because of symmetry.
For a spherical bottom trap, the model further predicts the confinement of the translational modes on the horizontal $x,y$ directions. The frequencies $f_x$ and $f_y$ of these modes are determined by the geometrical shape of the trap (radius of the spherical bottom). 

Experimentally, these predictions are confirmed only in part. We typically observe 5 normal modes of the levitated magnet out of 6, as our detection system is blind to the motion along $\gamma$. An example of a spectrum showing the 5 modes is shown in Supplemental Material (Fig.\,S3). For 4 out of 5 observed modes the theoretical predictions are well-reproduced. The horizontal frequencies $f_x$ and $f_y$ are reproducibly around $10$\,Hz, as expected, independently of the magnet size. The frequencies $f_z$ and $f_\beta$ depend on the magnet size and are always in good quantitative agreement with the model. Indeed, by measuring $f_z$ and $f_\beta$ and inverting the model we can obtain the magnet radius and magnetic moment, and the radius is found to be consistent with independent optical determination. Details are reported in the Supplemental Material \cite{ahrens2025,Supplementary}.

The fifth mode that we observe can be identified as the $\alpha$ libration. We have multiple evidences that this is the case. For instance, we can excite the motion of this mode until it turns into a continuous rotation, and the detection coupling is maximized using an 8-shaped pick-up coil (which instead tends to suppress other modes for symmetry). However, the $\alpha$ confinement is not predicted by the model. We suspect that it is caused by a residual magnetic field in the trap. Note that for the validity of the model, a full understanding of the trapping details is not relevant as long as we can demonstrate that it is magnetic dipole trapping. For instance, the equilibrium orientation could be slightly tilted from the horizontal plane (as expected if the residual field features a vertical component) without affecting the conclusions of the experiment. However, if the $\alpha$ trapping was due to a nonmagnetic mechanism, e.g. electrical, or to magnetic quadrupole effects, the geometry depicted in Fig.\ref{fig1}a with the librational modes defined by the direction of $\bm \mu$ may be incorrect, invalidating our model.

In Ref. \cite{ahrens2025} using the same setup and the same type of levitated magnet, it was shown that the frequency $f_\alpha$ can be efficiently tuned by superposing an additional static horizontal magnetic field $\bm{B_a}$. The field can be applied through two 8-shaped coils similar to PK1, generating fields along two orthogonal directions. The measurements showed consistency with a model in which $\alpha$ is confined by the total field $\bm{B}=\bm{B_0}+\bm{B_a}$ where $\bm{B_0}$ is a residual magnetic field. According to this model, $\bm \mu$ aligns along the total field $\bm B$ and the corresponding potential $U=-\bm {\mu \cdot B}$ implies a trapping frequency:
\begin{equation}
f_\alpha^2 = \frac{1}{4 \pi^2} \frac{\mu B}{I}  \label{alignment}
\end{equation}

\begin{figure}[!ht]
\includegraphics[width=8.6cm]{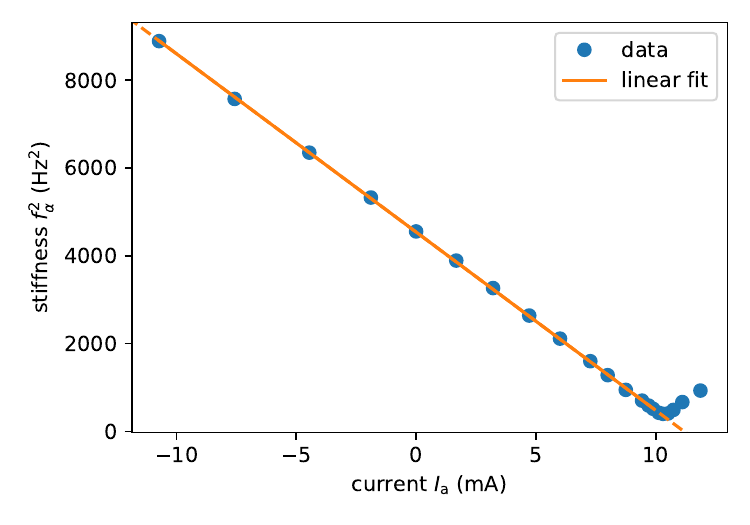} 
\caption{Square of the frequency $f_\alpha^2$ as a function of the current $I_\mathrm{a}$ in one of the two coils that we have used to apply a static magnetic field $\propto I_\mathrm{a}$ at the magnet location along two orthogonal directions on the horizontal plane. The current in the other coil was preliminary tuned to minimize $f_\alpha$, so that the field component orthogonal to $\bm \mu$ was approximately compensated. If the $\alpha$ confinement is determined by dipolar trapping in a static field with fixed direction, the angular stiffness $f_\alpha^2$ is expected to depend linearly with the total applied field $B=B_0+B_a$ according to Eq.\,(\ref{alignment}). The solid line is a linear fit over the wide linear region.}   \label{frequencytuning}
\end{figure}

To investigate potential limits of this model, we have performed additional measurements. In Fig.\,\ref{frequencytuning} we show similar tuning of $f_\alpha$ by applying an additional field $\bm {B_a}$ with the two coils. In particular, we show that, by a proper combination of the applied field components, we can almost completely compensate for the $\alpha$ mode trapping, reducing the angular stiffness, proportional to $f_\alpha^2$, by more than one order of magnitude.

In Fig.\,\ref{frequencytuning}, $f_\alpha$ was preliminarily minimized by tuning the field from one coil. In the residual field scenario, this implies that the component of the residual field orthogonal to $\bm \mu$ was compensated for. By varying the other (parallel) component, we expect a linear dependence of the angular stiffness, according to Eq.\,(\ref{alignment}). The linear dependence of $f_\alpha^2$ on a wide range of applied fields in Fig.\,\ref{frequencytuning} is thus a confirmation of this model. More importantly, it confirms that the dominant trapping mechanism is dipolar. The fact that we were not able to fully compensate the angular trapping suggests that an additional weaker non-dipolar trapping mechanism may be present. This could be a quadrupolar effect, associated with a physical quadrupole moment or with geometrical asymmetries of the trap. For the purpose of this paper, it is important to note that any additional trapping is at most a few percent of the dominant dipolar trapping. In the worst case scenario, with a nondipolar trapping in a direction orthogonal to $\bm \mu$, the added stiffness would add quadratically, leading to a negligible change of direction of $\bm \mu$ at equilibrium. This justifies the model in Fig.\,\ref{fig1}, where $\alpha$ and $\beta$ are defined with respect to $\bm \mu$.

Notably, while the data shown in Fig.\,\ref{frequencytuning} point to a dominant dipolar trapping, presumably magnetic, an electric dipole in a residual electric field might in principle provide a similar intrinsic confinement. To rule out electrical trapping, we have converted the field coil used in the measurements of Fig.\,\ref{frequencytuning} into an electrode (by disconnecting the ground return line) and applied dc voltages up to $30$ V, corresponding to electric fields of about $10^3$ V/m at the magnet location. The frequencies of the librational modes are almost completely unaffected, with a relative frequency shift $\delta f/f  \sim 10^{-6}$ V$^{-1}$ for both modes. This rules out electrical trapping as a possible explanation of the $\alpha$ confinement. Instead, we can easily drive the magnet normal modes via the electrode, with higher efficiency in the translational modes. This is entirely expected because the magnet will certainly hold some residual charge, as in any other levitated system \cite{Dania2024}.

Having established that the $\alpha$ confinement is indeed dominated by a dipolar magnetic trapping, let us discuss the possible origin of the residual magnetic field field. The superconducting trap is made of Pb and is completely closed, so under ideal Meissner effect, it should completely shield external fields. Moreover, Pb is a type-I material that does not support flux penetration via vortices. However, it could feature mixed-state with normal regions, where magnetic flux could be trapped. If this were the case, one may expect a significant impact on the $z$ and $\beta$ modes, which are instead in good agreement with a model based on ideal Meissner behavior. 
We believe that a more likely explanation is given by flux trapped in the superconducting pick-up coil circuits (which are not simply connected) or flux vortices trapped in the wire structure of the pick-up or actuation coils (which are made of NbTi, a type II superconducting material supporting vortices). Flux trapping could arise because the circuits become superconducting at a temperature higher than Pb, when the trap is still normal conducting.

Note that despite the residual field may have a vertical component, we still expect the equilibrium orientation of $\bm \mu$ to be approximately horizontal, as shown in Figs.\,\ref{fig1} and \ref{fig3}, for two reasons. First, the $\beta$ confinement, which forces the magnet onto the horizontal plane, is much stronger than the $\alpha$ one. In fact, $f_\beta$ is typically at least $4-5$ times higher than $f_\alpha$, leading to an angular stiffness at least $\sim 20$ times higher. Thus, tilts from the horizontal plane are expected to be small. Second, if the residual field arises from flux trapped in the pick-up and actuation coils, while the trap behaves according to the Meissner effect, the residual field is expected to be bent by the Meissner screening currents so as to lie approximately horizontally at the magnet location. 

\end{appendices}

\end{document}